# STAR FORMATION IN THIN DISKS OF SPIRAL GALAXIES SEEN FACE-ON

V. E. Karachentseva,[1]. I. D. Karachentsev,[2] and O. G. Kashibadze[2]

*Estimates of the integrated (SFR) and specific (sSFR) rates of star formation are given for 181 galaxies of later Sc, Scd, and Sd types seen almost face-on. Their SFRs were determined from FUV fluxes in the GALEX survey. The median values of the sSFR are: -10.66 dex for Sc, -10.44 dex for Scd, and -10.40 dex for Sd types in units of $yr^{-1}$. The average value of the sSFR for these galaxies falls off smoothly from low-mass to giant disks. After accounting for photometric errors, the specific star formation rate has a small cosmic variation of 0.16 dex. In order to reproduce the observed stellar mass on a cosmic time of 13.8 Gyr, the galaxies without bulges viewed face-on must have had an SFR two-three times higher in the past than observed now.*

## 1. Introduction

The survey of the sky in the FUV band of the far ultraviolet ($\lambda_{ef} = 1539\text{Å}$, $FWHM = 269\text{Å}$) on the GALEX space telescope [1,2] has led to the detection of FUV fluxes in tens of thousands of galaxies. This has made it possible

───────

(1) Main Astronomical Observatory, National Academy of Sciences of Ukraine
(2) Special Astrophysical Observatory, Russian Academy of Sciences (SAO RAN), Nizhnii Arkhiz, Karachaevo-Cherkessian Republic, Russia; e-mail: ikar@sao.ru

to determine simultaneously a current rate of star formation in galaxies over a characteristic time scale of ~$10^8$ yr, corresponding to the average age of young blue stars. The specific rate of star formation, sSFR = SFR/$M^*$, per unit stellar mass increases monotonically along the Hubble sequence from early types S0, Sa ($T$ = 0, 1 along the Vaucouleurs scale) to late types Sm, Im, Ir ($T$ = 8, 9, 10) [3.4]. The main reason for the increase in sSFR with increasing morphological type $T$ is a reduction in the fraction of the old star population, concentrated mainly in a spheroidal subsystem (bulge). The behavior of the sSFR as a function of morphological type or of stellar mass in galaxies without visible signs of a bulge is still unclear. Melnyk et al. [5] have examined a sample of ~500 ultra-flat (UF) spiral edge-on galaxies [6] for which the contribution of a bulge to the overall brightness of the galaxy is negligible. They concluded that on going from Sc to Sd types the specific rate of star formation increases by a factor of two. This conclusion, however, needs independent confirmation, since in UF galaxies visible from the edge there is a strong internal extinction effect that is hard to take into account quantitatively.

A sample of 220 spiral galaxies without bulges that are visible almost strictly edge-on has been set up in [7]. It included galaxies of morphological types Sc, Scd, and Sd ($T$ = 5–7) with an angular diameter $a > 0'.8$, axis ratio log($a/b$)<0.05, and radial velocity relative to the center of the Local group $V_{LG}$<10000 km/s in the declination zone DEC>–30°. Internal extinction is small in these objects and essentially has no effect on the shape of the sSFR(T) or sSFR(M*) relationships. The important feature of these galaxies is their localization in regions of low cosmic density [8], where there are no noticeable traces of interactions with nearby neighbors.

In the following we estimate the integrated (SFR) and specific (sSFR) star formation rates for galaxies without bulges that are visible face-on and examine the dependence of these parameters on the stellar and hydrogen masses of the galaxies.

## 2. Initial observational data

Of the 220 Sc, Scd, and Sd galaxies in our sample that are seen face-on, 39 galaxies lay outside the area of the GALEX survey. The other galaxies were reliably detected in the FUV-band. Data on them are given in Table 1. It should be noted that in a number of cases the FUV-images of the galaxies are broken up into several fragments. This required an individualized approach to determining the integrated flux for each galaxy.

The columns of Table 1 contain: (1) the name of the galaxy; (2) the morphological type according to the Vaucouleurs scale which we determined from multicolor images in the PanSTARRS survey [9]; (3) the apparent total magnitude magnitude of the galaxy in the B-band from the HyperLEDA data base [10]; (4) the Galactic extinction [11]; (5) the magnitude of the galaxy in the FUV-band; (6) the visible magnitude in the 21 cm neutral hydrogen line [10]; (7) the integrated K-band magnitude of the galaxy calculated from the B-magnitude and the morphological type using the formula

$$K_B = B_t - A_G + T/4 - 4.60, \tag{1}$$

TABLE 1. Initial Data for *Sc-Sd* Galaxies Seen Face-On

| Galaxy | $T$ | $B_t$ | $A_G$ | $m_{FUV}$ | $m_{21}$ | $K_B$ | $D$ | $\log L_K$ | $\log M_{HI}$ | logSFR | $\log SFR/L_K$ |
|---|---|---|---|---|---|---|---|---|---|---|---|
| 1 | 2 | 3 | 4 | 5 | 6 | 7 | 8 | 9 | 10 | 11 | 12 |
| NGC7816 | 5 | 13.91 | 0.27 | 16.31 | 14.58 | 10.29 | 74 | 10.93 | 10.24 | 0.204 | -10.73 |
| UGC00044 | 7 | 16.70 | 0.15 | 18.70 | 17.80 | 13.70 | 89 | 9.73 | 9.11 | -0.685 | -10.41 |
| UGC00048 | 6 | 15.60 | 0.58 | 16.91 | 14.89 | 11.92 | 63 | 10.14 | 9.97 | 0.061 | -10.08 |
| NGC7834 | 6 | 15.35 | 0.63 | 17.22 | 15.96 | 11.63 | 74 | 10.40 | 9.68 | 0.112 | -10.29 |
| NGC0039 | 5 | 14.34 | 0.35 | 16.69 | 16.02 | 10.64 | 70 | 10.74 | 9.61 | 0.061 | -10.68 |
| UGC00160 | 6 | 16.50 | 0.21 | 17.56 | 15.97 | 13.19 | 68 | 9.70 | 9.61 | -0.416 | -10.12 |
| PGC1075005 | 6 | 15.57 | 0.15 | 17.18 | ... | 12.32 | 46 | 9.71 | ... | -0.654 | -10.36 |
| PGC002257 | 5 | 15.41 | 0.11 | 17.34 | ... | 11.95 | 59 | 10.07 | ... | -0.528 | -10.60 |
| IC1562 | 5 | 13.60 | 0.05 | 15.24 | 15.02 | 10.20 | 52 | 10.66 | 9.76 | 0.156 | -10.51 |
| NGC0198 | 5 | 13.10 | 0.11 | 16.12 | 16.16 | 9.64 | 74 | 11.19 | 9.60 | 0.155 | -11.04 |
| IC0043 | 5 | 13.95 | 0.29 | 16.70 | 15.40 | 10.31 | 70 | 10.88 | 9.86 | 0.014 | -10.86 |
| NGC0236 | 5 | 14.43 | 0.11 | 16.08 | 16.02 | 10.97 | 80 | 10.73 | 9.73 | 0.235 | -10.49 |
| NGC0255 | 5 | 12.41 | 0.15 | 14.26 | 13.51 | 8.91 | 21 | 10.39 | 9.57 | -0.166 | -10.56 |
| IC0056 | 6 | 15.15 | 0.12 | 16.65 | 16.25 | 11.93 | 85 | 10.40 | 9.69 | 0.072 | -10.33 |
| UGC00626 | 5 | 14.97 | 0.21 | 16.96 | ... | 11.41 | 80 | 10.55 | ... | -0.035 | -10.59 |
| ESO412-013 | 5 | 15.00 | 0.09 | 16.68 | ... | 11.56 | 78 | 10.47 | ... | -0.037 | -10.51 |
| ESO542-004 | 7 | 15.00 | 0.06 | 16.97 | ... | 12.10 | 78 | 10.26 | ... | -0.182 | -10.44 |
| IC1666 | 6 | 14.35 | 0.27 | 17.03 | 16.23 | 10.98 | 70 | 10.61 | 9.53 | -0.134 | -10.74 |
| PGC005023 | 6 | 15.78 | 0.14 | 18.76 | 18.44 | 12.54 | 27 | 9.16 | 7.82 | -1.754 | -10.91 |
| UGC00929 | 6 | 14.82 | 0.15 | 16.43 | 16.23 | 11.57 | 104 | 10.72 | 9.87 | 0.358 | -10.36 |
| NGC0575 | 5 | 13.72 | 0.27 | 16.11 | 15.51 | 10.10 | 46 | 10.59 | 9.45 | -0.134 | -10.73 |
| NGC0628 | 5 | 9.80 | 0.28 | 12.20 | 11.56 | 6.17 | 10 | 10.84 | 9.71 | 0.118 | -10.73 |
| UGC01148 | 5 | 15.45 | 0.71 | 18.02 | 16.48 | 11.39 | 68 | 10.42 | 9.41 | -0.216 | -10.64 |
| ESO543-021 | 5 | 14.95 | 0.08 | 16.63 | ... | 11.52 | 79 | 10.50 | ... | -0.014 | -10.51 |
| UGC01347 | 5 | 13.50 | 0.32 | 15.97 | 15.65 | 9.83 | 79 | 11.17 | 9.86 | 0.431 | -10.74 |
| PGC007210 | 5 | 14.97 | 0.13 | 16.12 | 14.88 | 11.49 | 112 | 10.81 | 10.48 | 0.527 | -10.28 |
| UGC01478 | 5 | 14.56 | 0.47 | 17.21 | 16.18 | 10.74 | 69 | 10.69 | 9.53 | -0.066 | -10.76 |
| UGC01546 | 5 | 14.76 | 0.33 | 16.75 | 15.63 | 11.08 | 33 | 9.92 | 9.11 | -0.628 | -10.54 |
| PGC007942 | 6 | 15.00 | 0.12 | 16.83 | 16.50 | 11.78 | 73 | 10.32 | 9.46 | -0.137 | -10.46 |
| PGC008142 | 5 | 14.91 | 0.10 | 17.30 | ... | 11.46 | 110 | 10.81 | ... | 0.021 | -10.79 |
| UGC02043 | 5 | 14.96 | 0.52 | 17.88 | ... | 11.09 | 73 | 10.60 | ... | -0.243 | -10.85 |
| UGC02094 | 5 | 13.74 | 0.28 | 16.26 | 15.37 | 10.12 | 73 | 10.99 | 9.91 | 0.214 | -10.78 |
| ESO479-022 | 5 | 15.39 | 0.13 | 17.05 | ... | 11.91 | 99 | 10.54 | ... | 0.053 | -10.49 |
| UGC02174 | 5 | 15.09 | 0.57 | 17.11 | 14.86 | 11.17 | 72 | 10.56 | 10.10 | 0.091 | -10.47 |
| ESO546-011 | 5 | 14.52 | 0.15 | 16.27 | 15.23 | 11.02 | 62 | 10.49 | 9.82 | -0.031 | -10.52 |

TABLE 1. (Continued)

| 1 | 2 | 3 | 4 | 5 | 6 | 7 | 8 | 9 | 10 | 11 | 12 |
|---|---|---|---|---|---|---|---|---|---|---|---|
| NGC1067 | 5 | 14.55 | 0.82 | 17.57 | 16.12 | 10.38 | 64 | 10.77 | 9.49 | -0.005 | -10.78 |
| UGC02323 | 5 | 15.66 | 1.08 | 18.72 | 16.62 | 11.23 | 111 | 10.91 | 9.77 | 0.217 | -10.69 |
| PGC012008 | 5 | 15.00 | 0.21 | 17.33 | ... | 11.44 | 129 | 10.96 | ... | 0.232 | -10.72 |
| UGC02623 | 7 | 15.59 | 1.02 | 19.07 | 15.36 | 11.72 | 63 | 10.22 | 9.78 | -0.467 | -10.69 |
| UGC02692 | 5 | 14.07 | 0.34 | 16.06 | 15.86 | 10.38 | 87 | 11.04 | 9.87 | 0.496 | -10.54 |
| UGC02712 | 6 | 15.50 | 1.26 | 20.26 | 17.58 | 11.14 | 97 | 10.83 | 9.27 | -0.380 | -11.21 |
| UGC02721 | 7 | 16.40 | 0.58 | 18.45 | 16.67 | 12.97 | 89 | 10.02 | 9.56 | -0.253 | -10.28 |
| NGC1325A | 7 | 13.40 | 0.09 | 16.03 | 15.87 | 10.46 | 15 | 9.48 | 8.33 | -1.214 | -10.69 |
| ESO548-035 | 5 | 13.49 | 0.27 | 15.77 | 14.82 | 9.87 | 57 | 10.88 | 9.91 | 0.195 | -10.68 |
| NGC1376 | 5 | 12.85 | 0.20 | 15.02 | 14.62 | 9.30 | 57 | 11.10 | 9.99 | 0.438 | -10.66 |
| PGC013714 | 6 | 15.10 | 0.35 | 17.41 | ... | 11.65 | 59 | 10.19 | ... | -0.371 | -10.56 |
| IC0342 | 5 | 9.40 | 2.41 | 12.80 | 8.03 | 3.64 | 3 | 10.81 | 10.07 | 0.470 | -10.34 |
| UGC02859 | 5 | 16.00 | 1.93 | 19.88 | 15.29 | 10.72 | 77 | 10.79 | 9.99 | 0.084 | -10.71 |
| UGC03051 | 6 | 16.00 | 0.97 | 18.37 | 16.13 | 11.94 | 95 | 10.49 | 9.83 | 0.131 | -10.36 |
| NGC1599 | 5 | 14.10 | 0.17 | 15.38 | 15.33 | 10.58 | 54 | 10.54 | 9.66 | 0.225 | -10.32 |
| IC0391 | 5 | 12.98 | 0.56 | 15.04 | 14.69 | 9.07 | 24 | 10.44 | 9.21 | -0.046 | -10.49 |
| ESO552-047 | 5 | 15.26 | 0.44 | 17.59 | ... | 11.47 | 90 | 10.63 | ... | -0.007 | -10.64 |
| PGC017323 | 7 | 13.49 | 0.34 | 14.92 | 14.59 | 10.30 | 30 | 10.14 | 9.45 | 0.024 | -10.12 |
| PGC018031 | 5 | 14.84 | 1.09 | 18.30 | 16.69 | 10.40 | 96 | 11.11 | 9.62 | 0.262 | -10.85 |
| IC0441 | 5 | 14.40 | 1.18 | 17.16 | 15.16 | 9.87 | 30 | 10.32 | 9.22 | -0.224 | -10.54 |
| UGC03574 | 6 | 13.20 | 0.25 | 14.66 | 13.74 | 9.85 | 24 | 10.13 | 9.59 | -0.134 | -10.26 |
| UGC03701 | 6 | 14.80 | 0.26 | 16.59 | 15.29 | 11.44 | 42 | 9.98 | 9.46 | -0.409 | -10.39 |
| UGC03806 | 6 | 15.26 | 0.44 | 18.10 | 16.61 | 11.72 | 74 | 10.36 | 9.43 | -0.385 | -10.74 |
| UGC03825 | 5 | 15.06 | 0.37 | 16.97 | 16.21 | 11.34 | 114 | 10.89 | 9.96 | 0.394 | -10.50 |
| UGC03886 | 5 | 16.00 | 0.27 | 16.57 | 15.67 | 12.38 | 69 | 10.04 | 9.74 | 0.041 | -10.00 |
| UGC03924 | 7 | 15.66 | 0.13 | 17.54 | 16.80 | 12.68 | 69 | 9.92 | 9.29 | -0.457 | -10.37 |
| UGC04074 | 5 | 13.75 | 0.18 | 16.54 | ... | 10.22 | 99 | 11.21 | ... | 0.295 | -10.92 |
| NGC2500 | 7 | 12.22 | 0.17 | 13.80 | 13.95 | 9.20 | 12 | 9.79 | 8.91 | -0.450 | -10.24 |
| NGC2514 | 5 | 14.01 | 0.15 | 16.17 | 15.65 | 10.51 | 65 | 10.73 | 9.70 | 0.051 | -10.68 |
| PGC086610 | 7 | 16.40 | 0.20 | 18.14 | 17.45 | 13.35 | 65 | 9.60 | 8.98 | -0.696 | -10.29 |
| PGC023378 | 6 | 14.49 | 0.21 | 17.19 | 15.59 | 11.18 | 60 | 10.39 | 9.65 | -0.380 | -10.77 |
| UGC04380 | 5 | 15.05 | 0.26 | 17.39 | 16.43 | 11.44 | 104 | 10.77 | 9.79 | 0.058 | -10.71 |
| UGC04445 | 5 | 15.06 | 0.29 | 17.11 | 15.60 | 11.42 | 88 | 10.63 | 9.98 | 0.044 | -10.59 |
| IC0509 | 5 | 13.83 | 0.16 | 15.87 | 15.49 | 10.33 | 74 | 10.92 | 9.87 | 0.290 | -10.63 |
| NGC2607 | 5 | 14.95 | 0.18 | 16.09 | 15.02 | 11.42 | 47 | 10.09 | 9.67 | -0.173 | -10.26 |
| UGC04536 | 5 | 15.42 | 0.41 | 18.06 | ... | 11.66 | 103 | 10.67 | ... | -0.101 | -10.77 |
| NGC2661 | 6 | 13.86 | 0.13 | 15.30 | 15.15 | 10.63 | 54 | 10.52 | 9.73 | 0.226 | -10.30 |

TABLE 1. (Continued)

| 1 | 2 | 3 | 4 | 5 | 6 | 7 | 8 | 9 | 10 | 11 | 12 |
|---|---|---|---|---|---|---|---|---|---|---|---|
| UGC04853 | 6 | 15.24 | 0.17 | 16.34 | 15.22 | 11.97 | 34 | 9.59 | 9.30 | -0.559 | -10.15 |
| PGC026687 | 6 | 14.72 | 0.13 | 16.40 | ... | 11.49 | 46 | 10.04 | ... | -0.353 | -10.39 |
| UGC05015 | 6 | 15.30 | 0.09 | 16.69 | 15.85 | 12.11 | 28 | 9.36 | 8.88 | -0.935 | -10.29 |
| UGC05153 | 5 | 15.65 | 0.13 | 16.82 | ... | 12.17 | 110 | 10.53 | ... | 0.236 | -10.29 |
| UGC05169 | 7 | 15.44 | 0.05 | 17.41 | ... | 12.54 | 106 | 10.34 | ... | -0.099 | -10.44 |
| NGC2967 | 5 | 12.28 | 0.44 | 15.26 | 13.46 | 8.49 | 31 | 10.89 | 9.93 | -0.002 | -10.90 |
| UGC05274 | 6 | 14.92 | 0.18 | 17.01 | 15.97 | 11.64 | 79 | 10.45 | 9.74 | -0.087 | -10.54 |
| ESO566-019 | 6 | 13.99 | 0.21 | 16.03 | 15.27 | 10.68 | 47 | 10.38 | 9.57 | -0.125 | -10.51 |
| ESO499-011 | 7 | 15.01 | 0.27 | 17.50 | 15.80 | 11.89 | 38 | 9.71 | 9.17 | -0.856 | -10.57 |
| PGC028556 | 5 | 14.84 | 0.19 | 17.47 | 16.07 | 11.30 | 98 | 10.77 | 9.89 | -0.077 | -10.85 |
| ESO567-010 | 6 | 14.44 | 0.29 | 17.09 | 16.43 | 11.05 | 38 | 10.05 | 8.92 | -0.672 | -10.72 |
| PGC029301 | 5 | 15.24 | 0.21 | 20.14 | ... | 11.68 | 125 | 10.83 | ... | -0.919 | -11.75 |
| UGC05474 | 6 | 14.89 | 0.10 | 16.16 | 15.37 | 11.69 | 80 | 10.44 | 9.99 | 0.197 | -10.24 |
| UGC05483 | 5 | 15.13 | 0.17 | 17.29 | 16.42 | 11.62 | 82 | 10.49 | 9.59 | -0.181 | -10.67 |
| PGC029882 | 5 | 14.54 | 0.23 | 17.19 | ... | 10.96 | 126 | 11.13 | ... | 0.283 | -10.84 |
| PGC029929 | 7 | 14.30 | 0.34 | 16.82 | 15.08 | 11.11 | 44 | 10.15 | 9.59 | -0.400 | -10.55 |
| NGC3184 | 6 | 10.41 | 0.07 | 12.53 | 12.89 | 7.24 | 11 | 10.50 | 9.26 | -0.092 | -10.59 |
| PGC030452 | 5 | 14.78 | 0.06 | 16.64 | ... | 11.37 | 87 | 10.64 | ... | 0.051 | -10.59 |
| PGC030830 | 5 | 14.91 | 0.08 | 17.01 | ... | 11.48 | 99 | 10.71 | ... | 0.030 | -10.68 |
| NGC3344 | 5 | 10.50 | 0.14 | 12.60 | 12.20 | 7.01 | 10 | 10.51 | 9.45 | -0.149 | -10.66 |
| PGC031979 | 7 | 14.05 | 0.16 | 14.97 | 14.61 | 11.04 | 31 | 9.88 | 9.47 | -0.103 | -9.98 |
| PGC032091 | 6 | 14.45 | 0.17 | 16.00 | 14.92 | 11.18 | 37 | 9.97 | 9.50 | -0.353 | -10.33 |
| NGC3433 | 5 | 13.30 | 0.15 | 15.14 | 14.26 | 9.80 | 35 | 10.48 | 9.71 | -0.069 | -10.55 |
| PGC032817 | 6 | 15.28 | 0.17 | 17.06 | ... | 12.01 | 110 | 10.59 | ... | 0.167 | -10.42 |
| NGC3506 | 5 | 13.16 | 0.13 | 15.73 | 15.34 | 9.68 | 86 | 11.31 | 10.06 | 0.458 | -10.85 |
| UGC06130 | 5 | 15.02 | 0.16 | 17.20 | 16.11 | 11.51 | 112 | 10.80 | 9.99 | 0.119 | -10.68 |
| UGC06194 | 6 | 14.62 | 0.07 | 15.92 | 15.43 | 11.45 | 35 | 9.82 | 9.25 | -0.444 | -10.26 |
| PGC034006 | 5 | 14.13 | 0.27 | 16.46 | 15.30 | 10.51 | 103 | 11.13 | 10.24 | 0.429 | -10.70 |
| UGC06335 | 6 | 14.91 | 0.02 | 16.64 | 15.26 | 11.79 | 41 | 9.82 | 9.45 | -0.634 | -10.45 |
| UGC06429 | 5 | 13.78 | 0.05 | 15.74 | 15.14 | 10.38 | 53 | 10.61 | 9.72 | -0.033 | -10.64 |
| IC0696 | 6 | 14.50 | 0.22 | 16.83 | 15.58 | 11.19 | 84 | 10.69 | 9.95 | 0.062 | -10.62 |
| UGC06528 | 5 | 14.12 | 0.05 | 16.19 | 16.31 | 10.72 | 46 | 10.35 | 9.13 | -0.336 | -10.68 |
| PGC036269 | 5 | 15.59 | 0.12 | 17.94 | 17.10 | 12.12 | 88 | 10.35 | 9.38 | -0.415 | -10.77 |
| PGC036353 | 5 | 14.14 | 0.32 | 16.51 | ... | 10.47 | 39 | 10.31 | ... | -0.394 | -10.70 |
| NGC3938 | 5 | 10.87 | 0.09 | 13.05 | 13.08 | 7.43 | 18 | 10.85 | 9.61 | 0.140 | -10.71 |
| ESO573-002 | 5 | 15.50 | 0.31 | 17.16 | 15.86 | 11.84 | 76 | 10.34 | 9.75 | -0.081 | -10.42 |
| NGC4136 | 5 | 11.70 | 0.08 | 13.70 | 13.70 | 8.27 | 17 | 10.46 | 9.31 | -0.178 | -10.64 |

TABLE 1. (Continued)

| 1 | 2 | 3 | 4 | 5 | 6 | 7 | 8 | 9 | 10 | 11 | 12 |
|---|---|---|---|---|---|---|---|---|---|---|---|
| NGC4195 | 6 | 15.29 | 0.08 | 16.54 | 15.82 | 12.11 | 61 | 10.04 | 9.57 | -0.205 | -10.24 |
| NGC4303 | 5 | 10.16 | 0.11 | 12.24 | 12.82 | 6.70 | 15 | 10.98 | 9.56 | 0.322 | -10.66 |
| NGC4303A | 6 | 13.53 | 0.16 | 14.82 | 14.61 | 10.28 | 15 | 9.55 | 8.84 | -0.677 | -10.23 |
| IC3267 | 5 | 14.12 | 0.09 | 16.96 | 16.79 | 10.68 | 38 | 10.20 | 8.77 | -0.774 | -10.97 |
| IC3271 | 5 | 14.57 | 0.14 | 16.81 | 16.17 | 11.08 | 97 | 10.85 | 9.84 | 0.137 | -10.72 |
| NGC4411B | 6 | 12.98 | 0.12 | 14.95 | 14.67 | 9.76 | 28 | 10.30 | 9.36 | -0.210 | -10.51 |
| NGC4535 | 5 | 10.56 | 0.13 | 13.46 | 12.91 | 7.08 | 16 | 10.89 | 9.57 | -0.099 | -10.98 |
| NGC4571 | 6 | 11.92 | 0.22 | 15.18 | 14.94 | 8.60 | 15 | 10.22 | 8.71 | -0.771 | -10.99 |
| NGC4653 | 5 | 12.77 | 0.14 | 14.86 | 14.30 | 9.28 | 40 | 10.80 | 9.81 | 0.150 | -10.65 |
| PGC042868 | 7 | 13.03 | 0.13 | 14.91 | 13.88 | 10.05 | 17 | 9.75 | 9.24 | -0.623 | -10.37 |
| ESO574-029 | 5 | 13.67 | 0.26 | 16.58 | ... | 10.06 | 83 | 11.13 | ... | 0.188 | -10.94 |
| NGC4688 | 7 | 12.60 | 0.10 | 14.80 | 13.60 | 9.65 | 7 | 9.14 | 8.58 | -1.371 | -10.51 |
| NGC4900 | 5 | 11.89 | 0.10 | 13.82 | 14.49 | 8.44 | 16 | 10.34 | 8.94 | -0.264 | -10.61 |
| UGC08153 | 6 | 14.49 | 0.13 | 16.07 | 15.24 | 11.26 | 38 | 9.97 | 9.39 | -0.387 | -10.35 |
| PGC045690 | 5 | 15.86 | 0.07 | 17.53 | ... | 12.44 | 67 | 9.99 | ... | -0.524 | -10.51 |
| NGC5068 | 6 | 10.50 | 0.44 | 12.80 | 12.36 | 6.96 | 5 | 9.92 | 8.78 | -0.605 | -10.53 |
| UGC08436 | 7 | 15.03 | 0.11 | 16.60 | 16.09 | 12.07 | 41 | 9.71 | 9.12 | -0.548 | -10.26 |
| NGC5154 | 5 | 14.73 | 0.07 | 16.35 | 15.61 | 11.31 | 77 | 10.56 | 9.86 | 0.064 | -10.49 |
| PGC048087 | 7 | 14.45 | 0.32 | 16.70 | 15.17 | 11.28 | 34 | 9.86 | 9.33 | -0.591 | -10.45 |
| ESO445-076 | 7 | 14.73 | 0.29 | 16.65 | 15.51 | 11.59 | 34 | 9.74 | 9.19 | -0.591 | -10.33 |
| UGC08877 | 7 | 15.18 | 0.02 | 17.37 | 16.82 | 12.31 | 37 | 9.52 | 8.74 | -1.014 | -10.54 |
| NGC5405 | 5 | 14.52 | 0.14 | 16.42 | 15.58 | 11.03 | 94 | 10.85 | 10.05 | 0.269 | -10.58 |
| PGC049982 | 5 | 16.00 | 0.09 | 17.62 | ... | 12.56 | 126 | 10.49 | ... | 0.004 | -10.48 |
| NGC5457 | 5 | 8.30 | 0.05 | 10.00 | 10.41 | 4.90 | 7 | 11.04 | 9.86 | 0.510 | -10.53 |
| UGC09008 | 6 | 15.29 | 0.09 | 17.19 | 16.34 | 12.10 | 73 | 10.20 | 9.52 | -0.298 | -10.50 |
| NGC5468 | 5 | 12.94 | 0.12 | 13.90 | 13.94 | 9.47 | 37 | 10.66 | 9.89 | 0.449 | -10.21 |
| NGC5476 | 6 | 13.34 | 0.14 | 15.71 | 14.86 | 10.10 | 34 | 10.34 | 9.45 | -0.330 | -10.67 |
| NGC5494 | 5 | 13.30 | 0.36 | 15.13 | 14.40 | 9.59 | 30 | 10.43 | 9.53 | -0.039 | -10.47 |
| UGC09144 | 6 | 15.95 | 0.06 | 17.56 | ... | 12.79 | 108 | 10.26 | ... | -0.134 | -10.39 |
| UGC09216 | 5 | 14.51 | 0.08 | 16.78 | 15.88 | 11.08 | 77 | 10.65 | 9.75 | -0.097 | -10.75 |
| NGC5660 | 5 | 12.38 | 0.10 | 14.22 | 14.34 | 8.93 | 36 | 10.85 | 9.71 | 0.284 | -10.57 |
| UGC09317 | 6 | 15.00 | 0.06 | 16.22 | 15.58 | 11.84 | 61 | 10.15 | 9.67 | -0.088 | -10.23 |
| ESO580-014 | 6 | 14.55 | 0.39 | 16.64 | 16.18 | 11.06 | 82 | 10.71 | 9.69 | 0.254 | -10.46 |
| UGC09837 | 5 | 13.81 | 0.09 | 15.47 | 15.10 | 10.37 | 39 | 10.35 | 9.47 | -0.154 | -10.50 |
| UGC09945 | 5 | 14.19 | 0.25 | 16.42 | 15.32 | 10.59 | 94 | 11.02 | 10.15 | 0.354 | -10.67 |
| NGC5989 | 6 | 13.56 | 0.10 | 16.02 | 16.24 | 10.36 | 42 | 10.41 | 9.08 | -0.305 | -10.72 |
| UGC10020 | 7 | 14.44 | 0.22 | 15.86 | 15.25 | 11.38 | 31 | 9.74 | 9.21 | -0.416 | -10.16 |

TABLE 1. (Concluded)

| 1 | 2 | 3 | 4 | 5 | 6 | 7 | 8 | 9 | 10 | 11 | 12 |
|---|---|---|---|---|---|---|---|---|---|---|---|
| PGC056010 | 6 | 15.42 | 0.13 | 17.48 | 17.46 | 12.19 | 63 | 10.03 | 8.94 | -0.514 | -10.55 |
| PGC056318 | 5 | 15.33 | 0.08 | 17.35 | ... | 11.90 | 82 | 10.38 | ... | -0.269 | -10.65 |
| PGC056639 | 6 | 15.77 | 0.10 | 20.37 | ... | 11.67 | 115 | 10.76 | ... | -0.480 | -11.24 |
| NGC6143 | 5 | 13.92 | 0.10 | 16.18 | 15.38 | 10.47 | 76 | 10.88 | 9.94 | 0.147 | -10.74 |
| PGC058201 | 5 | 15.69 | 0.23 | 17.38 | 15.76 | 12.11 | 117 | 10.60 | 10.16 | 0.139 | -10.46 |
| UGC10427 | 5 | 15.19 | 0.06 | 16.88 | 16.47 | 11.78 | 124 | 10.78 | 9.93 | 0.260 | -10.53 |
| IC1221 | 6 | 14.59 | 0.05 | 15.58 | 15.75 | 11.44 | 78 | 10.52 | 9.81 | 0.371 | -10.15 |
| UGC10590 | 6 | 14.06 | 0.13 | 16.42 | ... | 10.83 | 45 | 10.28 | ... | -0.385 | -10.67 |
| IC1236 | 5 | 14.23 | 0.35 | 17.02 | 16.22 | 10.53 | 85 | 10.96 | 9.70 | 0.097 | -10.86 |
| UGC10956 | 5 | 15.64 | 1.19 | 19.81 | ... | 11.10 | 94 | 10.82 | ... | -0.280 | -11.10 |
| NGC6493 | 6 | 15.44 | 0.17 | 16.63 | ... | 12.17 | 85 | 10.30 | ... | 0.115 | -10.19 |
| UGC11214 | 6 | 15.00 | 1.01 | 18.62 | 15.61 | 10.89 | 39 | 10.14 | 9.27 | -0.710 | -10.85 |
| NGC6711 | 5 | 13.71 | 0.37 | 16.72 | 16.44 | 9.99 | 68 | 10.98 | 9.42 | 0.045 | -10.94 |
| NGC6821 | 7 | 13.62 | 1.39 | 17.71 | 14.52 | 9.38 | 18 | 10.07 | 9.03 | -0.726 | -10.79 |
| NGC6946 | 5 | 9.60 | 1.48 | 13.20 | 10.00 | 4.77 | 8 | 11.21 | 10.14 | 0.446 | -10.76 |
| PGC902799 | 5 | 16.30 | 0.20 | 17.16 | 15.44 | 12.75 | 49 | 9.59 | 9.53 | -0.547 | -10.14 |
| NGC7137 | 5 | 13.05 | 0.60 | 15.89 | 15.32 | 9.10 | 23 | 10.39 | 8.93 | -0.392 | -10.79 |
| UGC11816 | 5 | 14.83 | 0.61 | 17.20 | 15.28 | 10.87 | 68 | 10.63 | 9.88 | 0.032 | -10.59 |
| ESO532-008 | 5 | 15.29 | 0.14 | 17.26 | ... | 11.80 | 87 | 10.47 | ... | -0.136 | -10.61 |
| PGC068549 | 5 | 14.23 | 0.43 | 17.01 | ... | 10.45 | 71 | 10.83 | ... | 0.011 | -10.82 |
| ESO602-027 | 5 | 14.48 | 0.19 | 16.62 | 16.06 | 10.95 | 80 | 10.74 | 9.71 | 0.081 | -10.66 |
| NGC7309 | 5 | 13.04 | 0.25 | 15.78 | 15.43 | 9.44 | 57 | 11.05 | 9.67 | 0.174 | -10.87 |
| ESO603-011 | 5 | 15.01 | 0.21 | 17.18 | ... | 11.45 | 114 | 10.85 | ... | 0.185 | -10.66 |
| UGC12192 | 7 | 16.50 | 0.53 | 18.24 | 16.44 | 13.13 | 93 | 10.00 | 9.69 | -0.175 | -10.17 |
| IC5261 | 6 | 13.88 | 0.19 | 15.84 | 15.47 | 10.59 | 46 | 10.40 | 9.47 | -0.086 | -10.48 |
| NGC7495 | 5 | 13.76 | 0.42 | 16.44 | 14.64 | 9.99 | 70 | 11.01 | 10.17 | 0.220 | -10.79 |
| NGC7535 | 6 | 14.28 | 0.29 | 16.63 | 15.34 | 10.89 | 67 | 10.61 | 9.85 | 0.005 | -10.60 |
| UGC12522 | 7 | 15.41 | 0.46 | 18.07 | 15.62 | 12.10 | 42 | 9.72 | 9.33 | -0.850 | -10.57 |
| UGC12585 | 7 | 14.63 | 0.31 | 16.64 | 15.52 | 11.47 | 53 | 10.17 | 9.57 | -0.188 | -10.36 |
| UGC12635 | 6 | 15.31 | 0.21 | 17.71 | 16.62 | 12.00 | 74 | 10.25 | 9.42 | -0.401 | -10.65 |
| PGC071751 | 5 | 14.74 | 0.28 | 17.47 | 16.33 | 11.11 | 115 | 10.99 | 9.92 | 0.128 | -10.86 |
| ESO605-016 | 5 | 13.23 | 0.14 | 15.99 | 15.70 | 9.74 | 110 | 11.50 | 10.13 | 0.575 | -10.92 |
| PGC072738 | 5 | 14.82 | 0.23 | 17.25 | 16.71 | 11.24 | 83 | 10.65 | 9.48 | -0.106 | -10.76 |
| UGC12838 | 5 | 14.86 | 0.19 | 16.73 | 16.18 | 11.32 | 101 | 10.79 | 9.87 | 0.241 | -10.55 |
| NGC7798 | 5 | 12.95 | 0.28 | 15.56 | 16.03 | 9.32 | 36 | 10.70 | 9.03 | -0.113 | -10.81 |

(this relationship was obtained by approximating the data shown in Fig. 19 of [12]); (8) the distance to the galaxy (in Mpc) determined from the radial velocity $V_{LG}$ for a Hubble parameter $H_0 = 73$ km/s/Mpc; for nearby galaxies with $V_{LG} < 2500$ km/s the distance is estimated using the model of [13], which accounts for a galaxy infall towards the Virgo cluster and the expansion of the Local cosmic void; for the closest galaxies the individual photometric distances from the data base of the SAO RAN for the Local volume galaxies, http://www.sao.ru/lv/lvgdb/ [14] are given; (9) the logarithm of the integrated K-band luminosity of the galaxy for an absolute solar magnitude of $K_\odot = 3^m.28$; (10) the logarithm of the hydrogen mass of the galaxy

$$\log M_{HI} = 12.33 - 0.4 m_{21} + 2 \log D \tag{2}$$

in units of the Sun's mass; (11) the logarithm of the integrated rate of star formation in $M_\odot$/yr, which is given [3] by

$$\log(\text{SFR}) = 2.78 - 0.4(m_{FUV} - 1.93 A_G) + 2 \log D; \tag{3}$$

and, (12) the logarithm of the specific rate of star formation in units of (yr)$^{-1}$ assuming that $M^*/L_K = 1 \times M_\odot/L_\odot$ [15].

The distribution of the spiral galaxies in our sample with respect to the integrated star formation rate and the K-luminosity is plotted in Fig. 1. The galaxies of types Sc, Scd, and Sd are shown, respectively, by squares, triangles, and circles. The galaxies in the graph are morphologically segregated: the maximum values of SFR $\sim 3 M_\odot$/yr correspond to massive Sc galaxies of type M101 and the minimum values of SFR $\leq 0.1 M_\odot$/yr are typical of dwarf *Scd* and *Sd* spirals.

Figure 2 reproduces the distribution of the galaxies with respect to integrated star formation rate and hydrogen mass. The value of the SFR in the disks of spirals of late types has a closer correlation with $M_{HI}$ than with $M^*$ with a slope of 0.86±0.04. This slope is considerably lower than the slope of 1.4±0.1 that describes the log(SFR) vs log$M_{HI}$ relationship for individual star-formation centers (the Schmidt-Kennicutt law [16]).

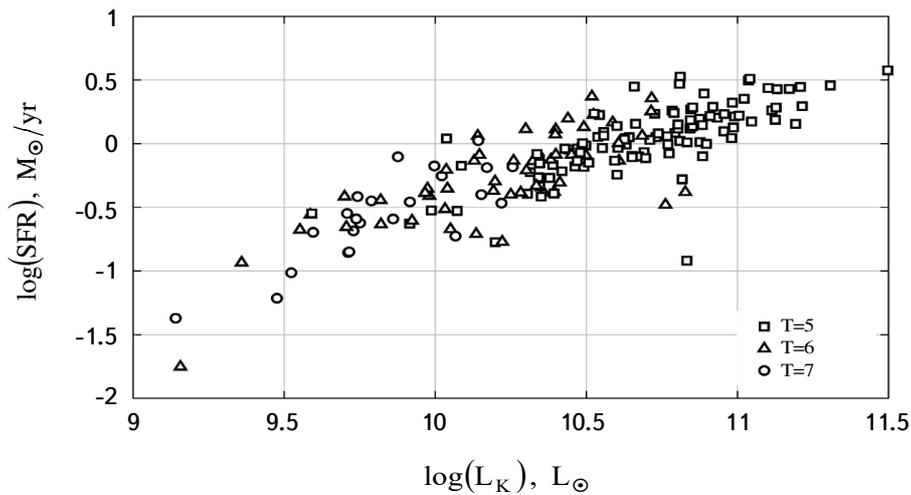

Fig. 1. Star formation rate and K-luminosity for galaxies of late types ($T = 5, 6, 7$) seen face-on.

The relationship between hydrogen mass and stellar mass (*K*-luminosity) for 146 galaxies from this sample is shown in Fig. 3. The notations for the morphological types are the same as in the previous figures. The three diagonal lines correspond to values of $M_{HI}/L_K$ equal to 1, 0.1, and 0.01. The median of this ratio is 0.15, while for the separate types of galaxies the median ratio is: 0.10 (Sc), 0.17 (Scd), and 0.28 (Sd). The data of Fig. 3 reflect the known fact that the relative content of neutral hydrogen in the disks of spiral galaxies falls off with increasing mass (luminosity) of a galaxy. This, in turn means that the process of conversion of the gas into a star was more intense in massive galaxies than in low-luminosity objects.

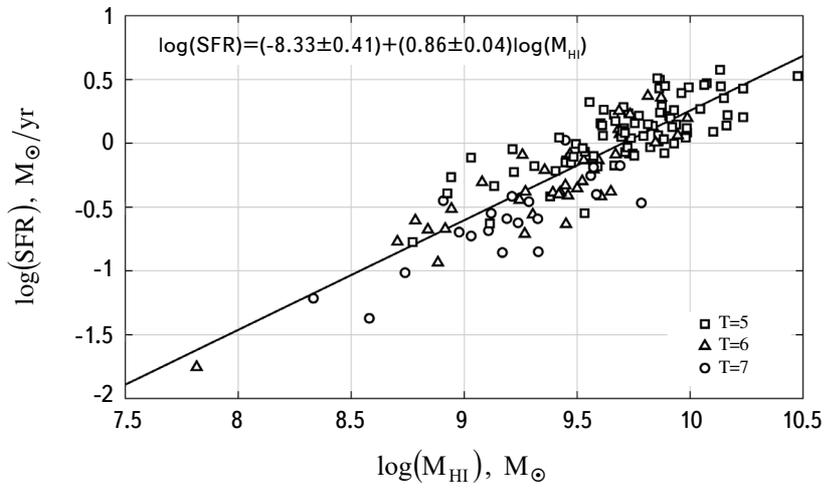

Fig. 2. Star formation rate and hydrogen mass for galaxies of late types oriented face-on.

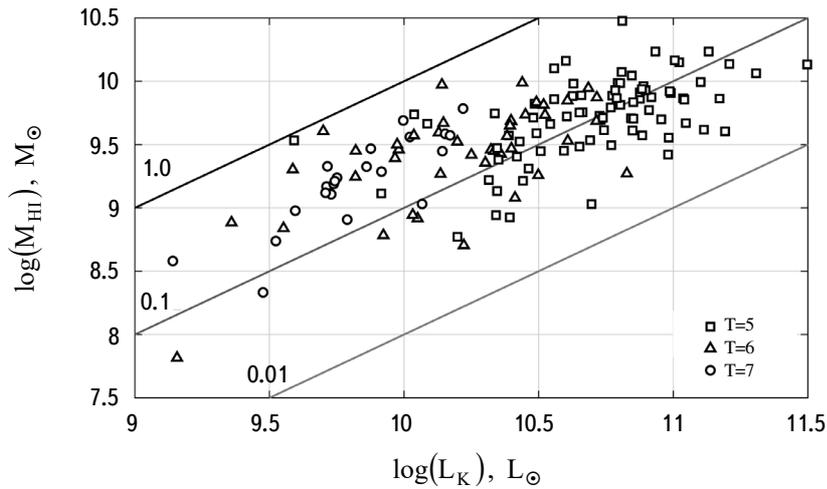

Fig. 3. Relationship of hydrogen mass and K-luminosity for galaxies of late types Sc, Scd, and Sd seen face-on.

TABLE 2. Rates of Star Formation Based on FUV and Hα Flux Data

| Galaxy | $D$, Mpc | $A_G$ | log[SFR(FUV)] | logF(Hα) | log[SFR(Hα)] |
|--------|----------|-------|---------------|----------|--------------|
| NGC628 | 10.2 | 0.28 | 0.12 | -10.84 | 0.22 |
| NGC2500 | 12.4 | 0.17 | -0.45 | -11.60 | -0.39 |
| NGC3184 | 11.1 | 0.07 | -0.09 | -11.12 | -0.04 |
| NGC3344 | 9.8 | 0.14 | -0.15 | -11.09 | -0.10 |
| NGC4136 | 17.0 | 0.08 | -0.18 | -11.82 | -0.36 |
| NGC4303 | 15.0 | 0.11 | 0.32 | -11.03 | 0.32 |
| NGC4688 | 7.0 | 0.10 | -1.37 | -11.90 | -1.21 |
| NGC4900 | 16.0 | 0.10 | -0.26 | -11.68 | -0.27 |
| NGC5068 | 5.1 | 0.44 | -0.60 | -10.83 | -0.34 |
| NGC5457 | 7.0 | 0.05 | 0.51 | -10.22 | 0.45 |
| NGC6946 | 7.7 | 1.48 | 0.45 | -10.47 | 0.63 |
| NGC7137 | 23.0 | 0.60 | -0.39 | -11.95 | -0.11 |
| NGC7798 | 36.0 | 0.28 | -0.11 | -12.13 | 0.03 |

Of the 181 galaxies in our sample, the star formation rate was determined from the Hα-flux, $F(H\alpha)$ in 13 cases. According to [16], the integrated star formation rate is given by

$$\log[\text{SFR}(H\alpha)] = \log F(H\alpha) - 0.244 A_G + 2\log D + 8.98, \qquad (4)$$

where the flux in the Hα line is measured in units of erg/cm²s and the distance $D$, in Mpc. Data on these galaxies from the NASA Extragalactic Database = NED (http://ned.ipac.caltech.edu) are given in Table 2. The estimates of the flux $F(H\alpha)$ for these galaxies were made by different authors on different telescopes. Nevertheless, the agreement in the values of the SFR based on the Hα and FUV fluxes is quite satisfactory. The average difference $\log[\text{SFR}(H\alpha)] - \log[\text{SFR}(FUV)]$ is $0.08 \pm 0.04$ with a dispersion of 0.12 for the difference. This comparison shows that the zero points of the empirical relations (3) and (4) used here are consistent with one another to within the typical errors in the flux measurements of ~0.1 dex.

## 3. Specific star formation rate in thin galactic disks

As noted above, thin spiral galaxies of late types without bulges are rather special objects whose evolution is little subject to external influence: bursts of star formation owing to tidal interaction of a massive neighbor or even the capture of a satellite during a merger. For this reason, these kinds of galaxies are extremely convenient for studying the autonomous evolution of spiral disks. The most important characteristic of the global evolution of a galaxy is the specific star formation rate relative to the unit stellar mass, i.e., $\text{sSFR} = \text{SFR}/M^*$. For estimating $M^*$ we have used the ratio $M^*/L_K \simeq 1 \times M_\odot/L_\odot$ [15], where the luminosity $L_K$ was determined from the apparent $B$-magnitude and average color index $\langle B-K \rangle$ which depends on the morphological type $T$.

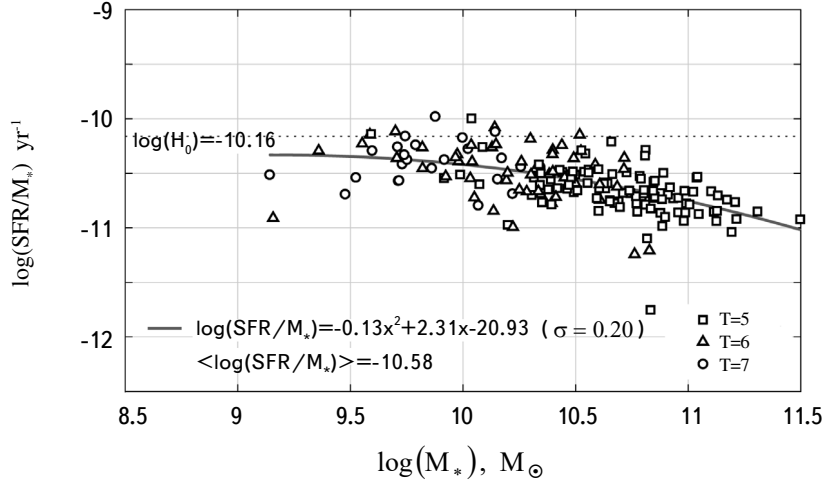

Fig. 4. The distribution of late-type galaxies observed face-on with respect to specific star-formation rate and stellar mass.

Figure 4 shows the distribution of the spiral disks of the galaxies without bulges oriented face-on with respect to the specific star formation rate and stellar mass. The dotted horizontal line corresponds the Hubble parameter $H_0 = (13.8 \times 10^9 \text{ yr})^{-1}$, when the galaxy is able to produce the observed mass at the observed rate SFR over a cosmological time of 13.8 billion years. As the quadratic regression curve (solid line) shows, the average specific SFR in thin disks decreases with increasing luminosity of the galaxy. These data show that in the past the average SFR in low-mass galaxies was a factor of two higher that that observed now, while for the most massive galaxies the earlier rate was a factor of 3-4 higher.

The spread of the galaxies relative to the regression fit equals 0.20 on a logarithmic scale. The average error in the K-luminosity owing to errors in measuring the B-magnitude ($0^m.25$) and the morphological type ($0^m.18$) is $0.^m31$. Making a quadratic subtraction of the errors, we find the true (cosmic) variation in the SFR magnitude to be 0.16 dex. This means that the conversion of the gas into a star in the thin disks of spiral galaxies is a fairly uniform process.

It should be noted that the HyperLEDA data base [10] contains estimates of the integrated $K_t$ magnitude for most of the objects in our sample. Estimates of $K_t$ were obtained mainly from the 2MASS survey [12]. However, because of a short exposure time (~8 s), the 2MASS survey systematically undercounts the contribution of the peripheral regions in the integrated luminosity of the galaxies, especially for objects with a blue color. Figure 5 reproduces the relationship between the $K_B$ and $K_t$ magnitudes for 159 galaxies in the sample examined here. The regression line for them is shown as $K_t = 0.58 + 0.99 K_B$. The average difference in the stellar magnitudes is $\langle K_t - K_B \rangle = 0^m.49$ increasing toward the faintest galaxies. The mean square deviation from the regression line is equal to $0^m.59$, and this also increases from bright to faint galaxies. Since the mean square error in the $K_B$-magnitude is $0^m.31$ according to our estimate, the main error, $\sigma(K_t) = 0^m.50$, is attributable to the fraction of the independently measured $K_t$-magnitude.

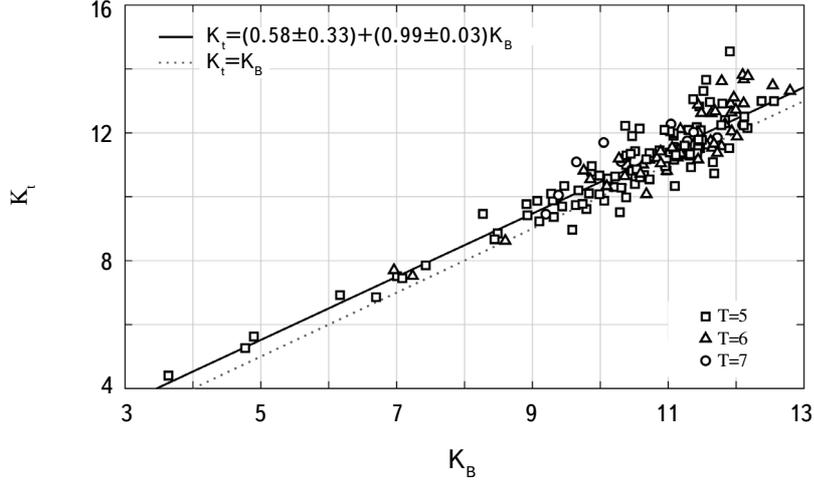

Fig. 5. Relationship between the integrated K-magnitude from HyperLEDA and the apparent K-magnitude derived from the color index (B-K) and the morphological type for Sc, Scd, and Sd galaxies seen face-on.

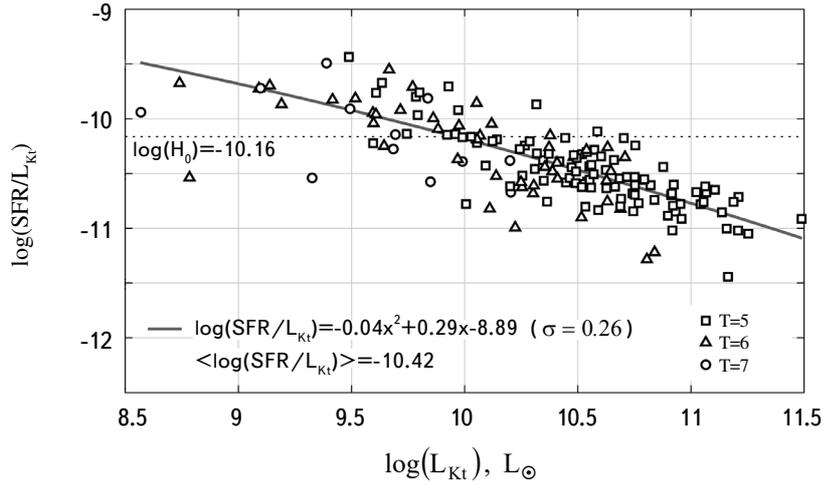

Fig. 6. The distribution of galaxies of late types seen face-on with respect to the specific star-formation rate and the $K$-luminosity based on HyperLEDA data.

In the photometric $K_t$ system, the relationship between the specific SFR and the $K_t$-luminosity takes the shape shown in Fig. 6. These data show that the transformation from $K_B$ to $K_t$ substantially increases the number of low-mass galaxies lying above the limit $sSFR = H_0$. The dispersion of the galaxies relative to the regression curve also increases, with $\sigma(\text{sSFR}) = 0.26$ dex. This value is consistent with the obtained estimate of $\sigma(K_t) = 0^m.50$ or $\sigma(L_K) = 0.20$ dex, if the true variation of the specific SFR equals 0.16 dex. This parameter shows that the formal use of the $K_t$-magnitude scale may lead to a systematic distortion in the picture of star formation in the disks of spiral galaxies, especially in their low-luminosity region.

## 4. The specific baryonic star formation rate

It is well known that when the luminosity a late type galaxy is lower, the fraction of its mass in the gaseous state will be larger. This behavior is evident for thin spiral disks, as well (see Fig. 3). It indicates that the conversion of the gas in a star for low-mass galaxies is less effective than in high-luminosity galaxies. In the course of the secular evolution the stellar mass of a galaxy increases, but the total baryon mass of the stars and the gas remain constant if the galaxy is isolated and is not supplemented with intergalactic gas. In this case it is possible to introduce the concept of the specific baryonic star formation rate sbSFR=SFR/$M_{bar}$, where

$$M_{bar} = M^* + M_{gas} = \Upsilon_K^* L_K + \eta M_{HI} . \qquad (5)$$

Until now we have assumed that $\Upsilon_K^* = 1$ [15]. However, McGaugh and Schombert [17] have advanced weighty arguments in favor of a smaller value, $\Upsilon_K^* = 0.6$. According to [18], another parameter $\eta$, equal to 1.85, accounts for the contribution of helium and molecular hydrogen to the total mass of the gas. In fact, the value of $\eta$ is still not fully known, since the fraction of molecular gas is not constant, but falls off toward low-mass galaxies. On the other hand, there are indications that galactic disks can be surrounded by significant masses of hot ionized

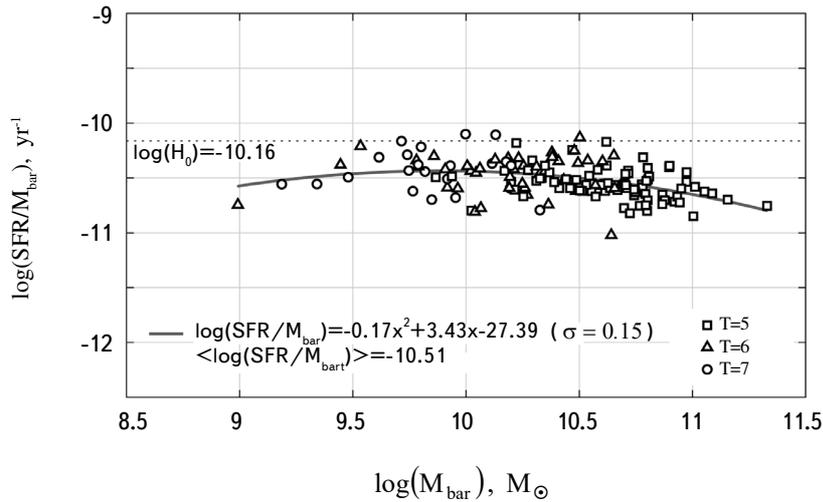

Fig. 7. The distribution of Sc, Scd, and Sd-type galaxies seen face-on with respect to the specific baryon rate of star formation and the baryon mass.

gas.

Assuming that $M_{bar} = 0.6 L_K + 1.85 M_{HI}$, we have calculated the specific baryonic star formation rate for the galaxies in our sample. The results are shown in Fig. 7. These data show that the dispersion of the galaxies relative to the regression curve decreases to the level of the cosmic variation, $\sigma(\text{sbSFR}) = 0.15$ dex. To within the photometric errors, all the galaxies of types Sc, Scd, and Sd lie under the horizontal line $\text{sbSFR} = H_0$, thereby showing that the contemporary star-formation rates for these galaxies are not enough to convert the initial mass of gas in the star over the cosmological time $H_0^{-1}$.

## 5. Concluding comments

We have examined a sample of 220 spiral galaxies without bulges that are oriented face-on. The integral and specific star formation rates for 181 of them have been determined using FUV fluxes from the GALEX survey. Of these, 106 are type Sc, 51 are type Scd, and 24 are type Sd. The resulting median values of the specific SFR for them are -10.66 dex, -10.33 dex, and -10.40 dex in units of yr$^{-1}$. Over the cosmological time of 13.8 Gyr, at the current rate of star formation, a typical spiral late-type galaxy can produce only 1/3-1/2 of its observed stellar mass. Thus, in the past their average SFR must have been 2-3 times greater than observed now. It should be noted that for a factor of $\Upsilon_K^* = 0.6$, rather than the factor of 1.0 we used in Eq. (5), the difference between the past and current SFR becomes a factor of 1.7 smaller.

The contemporary specific SFR is lower when the stellar mass of a spiral galaxy is higher. However, for more massive Sc-galaxies the observed ratio of the mass of the gas to the stellar mass (0.10) is considerably smaller than in Scd (0.17) and Sd (0.28) galaxies. Therefore, the efficiency of the process for converting the gas into the stars is higher when the mass of a spiral disk is higher.

With the contribution of photometric errors eliminated, we found a value of 0.16 dex for the dispersion of the sSFR for a fixed mass. The small scatter in the galaxies relative to the regression curve $\text{sSFR}(M^*)$ indicates a uniformity of the star formation process in the disks of thin spiral galaxies located in regions with low cosmic densities.

The star formation rate and the relative hydrogen content in the disks of Sc, Scd, and Sd galaxies seen edge-on [5,19] is in good agreement with the parameters obtained in this paper for spiral galaxies without bulges seen face-on. This agreement, in particular, implies that thin disks of Sc-Sd galaxies are essentially transparent in the 21 cm line for arbitrary angles of inclination of the disk to the line of sight.

This work was supported by RNF grant No. 19-12-00145. The HyperLEDA and NED data bases were used in this work, as well as data from the GALEX satellite and the PanSTARRS sky survey.

# REFERENCES

1.  D. C. Martin, J. Fanson, D. Schiminovich, et al., Astrophys. J. **619**, L1 (2005).
2.  A. Gil de Paz, S. Boissier, B. F. Madore, et al., Astrophys. J. Suppl **173**, 185 (2007).
3.  J. C. Lee, A. Gil de Paz, C. Tremonti, et al., Astrophys. J. **706**, 599 (2009).
4.  I. D. Karachentsev and E. I. Kaisina, Astron. J. **146**, 46 (2013).
5.  O. V. Melnyk, V. E. Karachentseva, and I. D. Karachentsev, AstBu **72**, 1 (2017).
6.  V. E. Karachentseva, Y. N. Kudrya, I. D. Karachentsev, et al., AstBu **71**, 1 (2016).
7.  I. D. Karachentsev and V. E. Karachentseva, Mon. Not. Roy. Astron. Soc. **485**, 1477 (2019).
8.  I. D. Karachentsev, V. E. Karachentseva, and Y. N. Kudrya, AstBu **71**, 129 (2016).
9.  K. C. Chambers, E. A. Magnier, N. Metcalfe, et al., arXiv:1612. 05560 (2016).
10. D. Makarov, P. Prugniel, N. Terekhova, et al., Astron. Astrophys. **A13**, 570A (2014).
11. D. J. Schlegel, D. P. Finkbeiner, and M. Davis, Astrophys. J **500**, 525 (1998).
12. T. H. Jarrett, T. Chester, R. Cutri, et al., Astron. J **125**, 525 (2003).
13. E. J. Shaya, R. B. Tully, Y. Hoffman, et al., Astron. J **850**, 207 (2017).
14. E. I. Kaisina, D. I. Makarov, I. D. Karachentsev, et al., AstBu **67**, 115 (2012).
15. E. F. Bell, D. H. McIntosh, N. Katz, et al., Astrophys. J. Suppl **149**, 289 (2003).
16. R. C. Kennicutt, Ann. Rev. Astron. Astrophys **36**, 189 (1998).
17. S. S. McGaugh and J. M. Schombert, Astron. J. **148**, 77 (2014).
18. M. Fukugita and P. J. E. Peebles, Astrophys. J **616**, 643 (2004).
19. S. S. Kaisin, I. D. Karachentsev, H. Hernandez-Toledo, et al., AstBu **73**, 1 (2020).